\documentclass{article}
\font \msb=msbm10 scaled \magstep1
\newcommand{\bR}{\mbox{\msb R} }
\newcommand{\bC}{\mbox{\msb C} }
\newcommand{\bP}{\mbox{\msb P} }

\def\a{{\alpha}}
\def\b{{\beta}}
\def\g{{\gamma}}
\def\d{{\delta}}

\def\l{{\lambda}}
\def\m{{\mu}}
\def\s{{\sigma}}
\def\D{{\Delta}}

\def\w{{\sf w }}

\newcommand{\sm}{\mbox{\bf s}}
\newcommand{\bs}{\mbox{\bf S}}
\newcommand{\bc}{\mbox{\bf C}}
\newcommand{\bu}{\mbox{\bf u}}
\newcommand{\p}{\mbox{\cal p}}
\newcommand{\q}{\mbox{\cal q}}

\def\B{{\cal B}}
\def\C{{\cal C}}
\def\D{{\cal D}}
\def\F{{\cal F}}
\def\H{{\cal H}}
\def\L{{\cal L}}

\def\U{{\cal U}}
\def\f{\mbox{\bf f}}
\def\e{\mbox{\bf e}}
\def\h{\mbox{\bf h}}
\def\te#1{{\widetilde{#1}}}
\def\pp{{\partial}}

\def\on#1#2{\mathop{\vbox{\ialign{##\crcr\noalign{\kern2pt}
$\scriptstyle{#2}$\crcr\noalign{\kern2pt\nointerlineskip}
\kern-2pt$\hfil\displaystyle{#1}\hfil$\crcr}}}\limits}
\def\nn{ \nonumber }
\def\bq{ \begin{equation} }
\def\eq{ \end{equation} }
\def\ben{ \begin{eqnarray} }
\def\en{ \end{eqnarray} }
\def\ll{ \label }
\def\frac#1#2{{#1\over #2}}
\def\dfrac#1#2{{\displaystyle{#1\over#2}}}

\newtheorem{prop}{Proposition}

\begin{document}

\title{
{\hfill\normalsize\sf solv-int/9811110}
\vskip1truecm
Duality between integrable St\"{a}ckel systems.}
\author{
 A.V. Tsiganov\\
{\small\it
 Department of Mathematical and Computational Physics,
 Institute of Physics,}\\
{\small\it St.Petersburg University, 198 904,  St.Petersburg,  Russia,}\\
{\small\it e-mail: tsiganov@mph.phys.spbu.ru}
}
\date{}
\maketitle

\vskip0.5cm
{
For the St\"{a}ckel family of the integrable systems a non-canonical
transformation of the time variable is considered. This transformation
may be associated to the ambiguity of the Abel map on the corresponding
hyperelliptic curve. For some St\"{a}ckel's systems with two degrees of
freedom the $2\times 2$ Lax representations and the dynamical $r$-matrix
algebras are constructed.  As an examples the Henon-Heiles systems,
integrable Holt potentials and the integrable deformations of the Kepler
problem are discussed in detail.}


\section{Introduction}
In her celebrated papers \cite{kow89} published in 1889 S.Kowalewski
discovered a necessary condition for an $n$-dimensional system to be
completely integrable. This criterion enabled her to classify all
integrable solid body motion about a fixed point and introduce the
separation variables for her celebrated top. Evidently, it was the first
application of singularity analysis to a concrete physical problem.
Recall, that the method of singularity analysis associates integrability
with the Kowalewski-Painlev\'{e} property, i.e. a movable polelike
singularity $(t-t_0)^{-m}$ in the solution of the equations of motion,
here space and time must be thought of as complex \cite{rgb89,avm89}.

There exist cases, however, of integrable hamiltonian system with
rational integrals of the motion, whose analytic structure permits
solutions with algebraic singularities of the type $(t-t_0)^{1/k}$, ($k$
being a positive integer larger than one). This led to the introduction
of the "weak" Painlev\'{e} property \cite{rgb89}. A simple change of the
independent variable, in and of itself, does not turn a "weak"
Painlev\'{e} system into one that satisfies the usual
Kowalewski-Painlev\'{e} criterion. In particular, the simple-minded idea
to take $(t-t_0)^{1/k}$ as a new independent variable does not lead to a
Painlev\'{e} expansion for all the solutions. According by \cite{rgb89},
any transformation that modifies the nature of the singular expansions
must also involve a change of the dependent variables in order to
reestablish the Painlev\'{e} property. Such transformations, however, if
they exist, are expected to be quite nontrivial and difficult to
generalize to other examples.

Of course, transformations, which use change of the independent time
variable $t$, are non-canonical transformations. The classical example of
the such non-canonical transformations is the  duality of  the Kepler
problem to the geodesic motion on the sphere \cite{mos70} or to the
harmonic oscillator \cite{gol80}.

Another important to us example is the Kolosoff transformations for the
Ko\-wa\-lew\-ski top \cite{kol01}. In this case the Ko\-wa\-lew\-ski
separation variables coincide to the standard elliptic coordinates
$\{q_1,q_2\}$ at the plane $(x,y)$ after the non-canonical change of the
time
\bq
d\te{t}=\Bigl(\,q_1(t)+q_2(t)\,\Bigr)\,dt\ll{tkow}\,.
\eq
Recently, the non-canonical transformations relate the Kowalewski top
with the geodesic motion on $SO(4)$ \cite{avm88} and with the Neumann
system on the sphere $S^2$ \cite{hh87}.

In this paper we consider the fortiori integrable systems with the known
separation variables. Since we shall not discuss singularity analysis
\cite{rgb89} and theory of algebraic completely integrable systems
\cite{avm89} in detail. Recall, the variable separation method permits
one to reduce an integration problem with several degrees of freedom to a
sequence of one-dimensional integration problems. The inverse problem of
obtaining various classes of completely integrable systems starting from
a set of separated one-dimensional problems was started in the lectures
by Jacobi. In framework of this approach, Liouville and St\"{a}ckel
introduced a family of the simple systems integrable in quadratures (a
Liouville family is a particular case of a St\"{a}ckel family).

Here for the St\"{a}ckel system we introduce non-canonical
transformations of the time variable associated to the ambiguity of the
Abel map on the hyperelliptic curve. All these transformations depend on
coordinates only and, therefore, they are closed to the Kolosoff
\cite{kol01} change of the time (\ref{tkow}). For the some St\"{a}ckel
systems we propose the Lax pairs and $r$-matrix algebras. As an examples
the Henon-Heiles systems, integrable Holt potentials and the integrable
deformations of the Kepler problem are discussed in detail.


\section{Duality between the St\"{a}ckel systems}
\setcounter{equation}{0}
Before proceeding father it is useful to recall the classical work of
St\"{a}ckel \cite{st95}. The system associated with the name of
St\"{a}ckel \cite{st95} is a holonomic system on the phase space
$\bR^{2n}$ equipped with the canonical variables $\{p_j,q_j\}_{j=1}^n$,
with the standard symplectic structure $\Omega_n$ and with the following
Poisson brackets
\bq \Omega_n=\sum_{j=1}^n dp_j\wedge dq_j\,,\qquad
\{p_j,q_k\}=\d_{jk}\,.\ll{stw}
\eq
The nondegenerate $n\times n$ St\"{a}ckel matrix $\bs$, whose $j$ column
$\sm_{kj}$ depends only on $q_j$
\[
\det \bs\neq 0\,,\qquad \dfrac{\pp \sm_{kj}}{\pp q_m}=0\,,
\quad j\neq m\]
defines $n$ functionally independent integrals of motion
\bq
I_k=\sum_{j=1}^n c_{jk}\left(p_j^2+U_j\right)\,,
\qquad c_{jk}=\dfrac{\bs_{kj}}{\det\bs}\,.
\ll{int}
\eq
which are quadratic in momenta. Here $\bc=[c_{ik}]$ denotes inverse
matrix to $\bs$ and  $\bs_{kj}$ be cofactor of the element $\sm_{kj}$.

Each integral $I_k$ (\ref{int}) may be associated to the time variable
$t_k$, such that for any function $\xi(\p,\q)$ one gets
\[\dfrac{d \xi(\p,\q)}{dt_k}=\{I_k,\xi(\p,\q)\}\,.\]
By definition the first integral $I_1=H$ be the Hamilton function
associated to the time $t$.  The common level surface of the integrals
(\ref{int})
\bq
M_\a=\left\{z\in \bR^{2n}: I_k(z)=\a_k\,,~k=1,\ldots,n\right\}\ll{alp}
\eq
is diffeomorphic to the $n$-dimensional real torus
and  one immediately gets
\bq
p_j^2=\left(\dfrac{\pp {\cal S}}{\pp q_j}\right)^2=
\sum_{k=1}^n \a_k\sm_{kj}(q_j)-U_j(q_j)\,,\ll{stc}
\eq
where ${\cal S}(q_1\,\ldots,q_n)$ is a reduced action function
\cite{arn89}. If this real torus is a part of complex algebraic torus,
then the corresponding mechanical system is called an algebraic
completely integrable system \cite{avm89}.

The corresponding Hamilton-Jacobi equation on $M_\a$
\bq
\dfrac{\pp {\cal S}}{\pp t}+
H(t,\dfrac{\pp {\cal S}}{\pp q_1},\ldots,
\dfrac{\pp {\cal S}}{\pp q_n},q_1,\ldots,q_n)=0\,,\qquad\Rightarrow\qquad
c_{j1}\,\dfrac{\pp {\cal S}}{\pp q_j}\,
\dfrac{\pp {\cal S}}{\pp q_j}=E\,,\ll{hjeq}
\eq
admits the variable separation
\bq
{\cal S}(q_1\,\ldots,q_n)=\sum_{j=1}^n {\cal S}_j(q_j)\,,\qquad {\cal
S}_j(q_j)=\int{\sqrt{F_j(q_j)~~}~d q_j}\,.\ll{stac}
\eq
Here the functions $F_j(\l)$ depend on $n$ parameters $\{\a_k\}_{k=1}^n$
\[
 F_j(\l)=\sum_{k=1}^n \a_k\sm_{kj}(\l)-U_j(\l)\,.
\]
Coordinates $q_j(t,\a_1,\ldots,\a_n)$ are determined from the equation
explicitly depending on time
\bq
\sum_{j=1}^n\int_{\g_{0}(p_0,q_0)}^{\g_j(p_j,q_j)}\dfrac{\sm_{1j}(\l) d\l}
{\sqrt{\sum_{k=1}^n \a_k\sm_{1j}(\l)-U_j(\l)}}=\b_1=t\,,\ll{stinv1}
\eq
and from other $n-1$ equations
\bq
\sum_{j=1}^n\int_{\g_{0}(p_0,q_0)}^{\g_j(p_j,q_j)}\dfrac{\sm_{kj}(\l) d\l}
{\sqrt{\sum_{k=1}^n \a_k\sm_{kj}(\l)-U_j(\l)}}=\b_k\,,
\qquad k=2,\ldots,n\,.\ll{stinv}
\eq
The solutions of the problem is thus reduced to solving a sequence of
one-dimensional problems, which is the essence of the method of
separation of variables.

Now we turn to the non-canonical change of the time and prove the
following
\begin{prop}
If the two St\"{a}ckel matrices $\bs$ and $\te{\bs}$ be distinguished the
first row only
\[\sm_{kj}=\te{\sm}_{kj}\,,\qquad k\neq 1\,,\]
the corresponding St\"{a}ckel systems with the following Hamilton
functions
\bq
\te{H}=v(\q)\,{H}\,,\qquad
v(\q)=\dfrac{\det{\bs}(q_1,\ldots,q_n)}{\det\te{\bs}(q_1,\ldots,q_n)}
\,,\ll{dham}
\eq
are related by non-canonical change of the time.
\end{prop}
In fact, the corresponding Hamilton functions $H$ and $\te{H}$ obey to
the equation (\ref{dham}), which follows from the definitions of the
hamiltonians
\bq
H=\sum_{j=1}^n c_{j1}\left(\,p_j^2+U_j(q_j)\,\right)
\ll{ham}
\eq
and entries of the inverse matrix
\[c_{j1}=\dfrac{\bs_{1j}}{\det\bs}=
\dfrac{1}{\det\bs}\dfrac{\pp \det\bs}{\pp \sm_{1j}}\,.
\]
Equation (\ref{dham}) defines an implicit change of the time $t\to\te{t}$
associated to the integrals $H=I_1$ and $\te{H}=\te{I_1}$, respectively.
On the other hand the equation (\ref{stinv1}) may be considered as an
explicit determination of this transformation $t\to\te{t}$. In contrast
with the general coupling constant metamorphosis discussed  in
\cite{rgb89} equation (\ref{dham}) is independent on the any constant
entering in the potential $U$.

Obviously, by using row by row transformations of the St\"{a}ckel
matrices with the associated $t_k\to\te{t}_k$ transformations we can
reduce the given St\"{a}ckel system to any other St\"{a}ckel system on
$\bR^{2n}$.

As an example, let us consider three matrices
\bq
\bs=\left({\begin{array}{cc} 1 & 1 \\ 1 & -1\end{array}}\right)\,,\qquad
\te{\bs}=\left({\begin{array}{cc} q_1 & q_2 \\ 1 & -1\end{array}}\right)\,,\qquad
\widehat{\bs}=\left({\begin{array}{cc} q_1^2 & q_2^2 \\ 1 & -1\end{array}}\right)
\,.\ll{sh}
\eq
The corresponding hamiltonians $H$, $\te{H}$ and $\widehat{H}$ defined
by (\ref{ham}) are dual (\ref{dham})
\ben
\te{H}&=&\dfrac{(q_1+q_2)^{-1}}2\,H\,,\nn\\
\ll{dham1}\\
\widehat{H}&=\,&\dfrac{(q_1^2+q_2^2)^{-1}}2\,H=
\dfrac{q_1+q_2}{q_1^2+q_2^2}\,\te{H}\,,\nn
\en
for any potentials $U$. For the hamiltonians $\te{H}$ and $\widehat{H}$
the change of the time (\ref{dham1}) is closed to the Kolosoff
transformation (\ref{tkow}) \cite{kol01} and for any function $\xi(\q)$
depending on coordinates only one gets
\bq
\dfrac{d\,\xi(\q)}{d\,\te{t}}=\{\te{H},\xi(\q)\}=
\dfrac{(q_1+q_2)^{-1}}2\,\{H,\xi(\q)\}=
\dfrac1{2\,(q_1+q_2)}\,\dfrac{d\,\xi(\q)}{d\,t}\,.\ll{tk}
\eq
For instance, let us consider uniform cubic potential
\bq
U(q_j)=2\alpha^2\,q_j^3+\b\,q_j^2+\g\,q_j+\delta\,,\ll{3pot}
\eq
which gives rise to the hamiltonian $H$
\bq H=\dfrac {1}{4}\,(p_1^2 +
p_2^2) +\a^2\,(q_1^3 +q_2^3) +\dfrac{\b}2\,(q_1^2+q_2^2)
+\dfrac{\g}2\,(q_1+q_2) +\d\,.\ll{hh1}
\eq
Using canonical transformation
\ben
q_1&=\dfrac{x+y}2\,,\quad p_1&=p_x+p_y\,,\nn\\
q_2&=\dfrac{x-y}2\,,\quad p_2&=p_x-p_y\,,\nn
\en
for the first system,  the more complicated transformation
\ben
q_1&=\dfrac34\,x^{2/3}+\dfrac{p_y}{3\,\a}\,,\quad
p_1&=p_x\,x^{1/3}-\dfrac{3\,\a}{2}\,y\,,\nn\\
q_2&=\dfrac34\,x^{2/3}-\dfrac{p_y}{3\,\a}\,,\quad
p_2&=p_x\,x^{1/3}+\dfrac{3\,\a}{2}\,y\,,\nn
\en
for the system associated to $\te{\bs}$ and the following change of
variables in the third case
\ben
q_1&=\sqrt{x} - \sqrt{y}\,,\quad
p_1&=p_x\,\sqrt{x}-p_y\,\sqrt{y}\,,\nn\\
q_2&=-i\,(\sqrt{x}+\sqrt{y})\,,\quad
p_2&=i\,(p_x\,\sqrt{x}+p_y\,\sqrt{x})\,,\nn
\en
one gets the Hamilton functions in the natural form
\ben
H&=&\dfrac12 \,(p_x^2 + p_y^2)
+\dfrac{\a^2}4\,x\,(x^2+3\,y^2)
+\dfrac{\b}4\,(x^2+y^2)
+\dfrac{\g}2\,x+\d\,,\nn\\
\nn\\
\te{H}&=&\dfrac12 \,(p_x^2 + p_y^2) +
\dfrac {9\,\a^2}{8}\,x^{-2/3}\,(\dfrac34\,x^{2}+y^2)
+\d\,x^{-2/3}+\dfrac{3\g}4\,,\quad{\mbox{\rm only
by}}\quad\b=0\,,\nn\\
\nn\\
\hat{H}&=&\dfrac12\,p_x\,p_y
-\dfrac{\b}2\,\dfrac{1}{\sqrt{x\,y}}
+\dfrac{\g}4\,\left(\dfrac{1+i}{\sqrt{x}}-\dfrac{1-i}{\sqrt{y}}\right)+\d\,,
\quad{\mbox{\rm by}}\quad \a=0\,.\nn
\en
under restriction $\b=0$ for the second case.

The system with the first Hamiltonian $H$ is so-called first integrable
case of the Henon-Heiles system \cite{rgb89}. The second Hamiltonian
$\te{H}$ is related to so-called Holt potential \cite{rgb89}. Note, the
second integral of motion is the polynomial of the third order in momenta
for the Holt system. The system with the third Hamiltonian $\widehat{H}$
may be considered as an integrable deformation of the Kepler problem.

Duality between the Henon-Heiles and the Holt  systems with the
Hamiltonians $H$ and $\te{H}$ may be considered as the known coupling
constant metamorphosis with respect to the constant $\g$ \cite{rgb89}.
The second known duality between the harmonic oscillator and the Kepler
problem with the Hamiltonians $H$ by $\a=0\,,\g=0$ and $\widehat{H}$ may
be considered as the coupling constant metamorphosis with respect to
another constant $\d$ \cite{rgb89,gol80}.

We can see that in practical circumstances the St\"{a}ckel approach is
not very useful because it is usually unknown which canonical
transformation have to be used in order to transform a Hamiltonian
(\ref{ham}) to the natural form $H=T+V$ \cite{gol80}. This problem was
partially solved for the uniform systems $U_j=U,~j=1,\ldots,n$ with
polynomial potentials by using the corresponding Lax pairs \cite{ts97d}.
Note, that the movable branch points of the type $(t-t_0)^{1/k}$ appear
in the expansions of the physical variables ($x,y$) after canonical
transformations.

Henceforth, we shall restrict our attention to the uniform St\"{a}ckel
systems, where all the polynomial potentials $U_j(q_j)= U(q_j)$ and
associated hyperelliptic curves $\C_j$ (\ref{stc}) are equal.


\section{Duality and Abel map.}
\setcounter{equation}{0}
Let us briefly recall some necessary facts about the
Abel map and the inverse Jacobi problem.

The set of point $\C(z,\l)$ satisfying
\bq
\C:\quad z^2=F(\l)=\sum_{k=0}^{2g+1}\,e_k\,\l^k
=\prod_{j=1}^{2g+1}(\l-\l_j)\,,\ll{curve}
\eq
is a model of a plane hyperelliptic curve of genus $g$. Here $F(\l)$
is polynomial without multiple zeros. Let us denote by ${\rm Div}(\C)$
the Abelian divisor group and denote by $J(\C)$ the Jacobian of the
curve $\C$. The Abel map puts into correspondence the point $D\in {\rm
Div}(\C)$ and the point $\bu\in J(\C)$ \cite{du81,bel97}
\bq \U:\quad {\rm Div}(\C) \to J(\C)\,,\ll{ajm}\eq
The set of all effective divisors $D=\g_1+\cdots+\g_n$ (the $\g_j's$
may be not mutually distinct) of ${\rm deg}\,n$ of $\C$ is called the
$n$th symmetric product of $\C$, and is denoted by $\C^{(n)}=S^n\C$.
The $C^{(n)}$ can be identified with the set of all unordered
$n$-tuples $\{\g_1,\ldots,\g_n\}$, where $\g_j$ are arbitrary elements
of $\C$. Now consider restriction of the Abel map (\ref{ajm}) to
$\C^{(n)}$
\bq \U:\quad \C^{(n)}\to J(\C)\,,\ll{rajm}\eq
where
\[\U(\g_1,\g_2\,,\ldots,\g_g)=\U(\g_1)+\U(\g_2)+\cdots+\U(\g_g)\,.\]
According to the Abel-Jacobi theorem this map is surjective and
generically injective if $n=g$ only \cite{du81,bel97}. If $n\neq g$ the
Abel map is either lack of uniqueness or degenerate. The corresponding
St\"{a}ckel system either has a dual system associated with the same
curve or it is a superintegrable system \cite{ts97d}.

Suppose that point $D=\g_1+\cdots+\g_k$, $k\leq g$ belongs to $\C^{(k)}$.
The differential of the Abel-Jacobi map (\ref{rajm}) at the point $D$ is
a linear mapping from the tangent space $T_D(\C^{(n)})$ of $\C^{(n)}$	 at
the point $D$ into the tangent space $T_{\U(D)}(J(\C))$ of $J(\C)$ at the
point $\U(D)$
\[\U_D^*:\quad
T_D(C^{(n)})\to T_{\U(D)}(J(\C))\,.\] Now suppose that $D$ is a generic
divisor, and $x_j$ is a local coordinate on $\C$ near the point $\g_j$.
Then $(x_1,\ldots,x_n)$ yields a local coordinate system near the point
$D$ in $\C^{(n)}$. Let $d\w_k$ ($k=1,\ldots,g$) is a basis for a space
$\H_1(\C)$ of holomorphic differentials on $\C$, and near $\g_j$
\bq
d\w_k=\phi_{kj}(x_j)dx_j\,,\ll{wd}
\eq
where $\phi_{kj}(x_j)$ is holomorphic.
It follows that the Abel-Jacobi map $\U$ can be expressed near $D$ as
\[
\U(z_1,\ldots,z_n)=\left(
\sum_{j=1}^n\int_{\g_0}^{x_j} \phi_{1j}(x_j)dx_j,\ldots,
\sum_{j=1}^n\int_{\g_0}^{x_j} \phi_{gj}(x_j)dx_j\right)\,.\]
Hence
\bq
\U_D^*=
\left(\begin{array}{ccc}
\phi_{11}(\g_1)&\cdots&\phi_{g1}(\g_1)\\
\vdots&\ddots&\vdots\\
\phi_{1k}(\g_n)&\cdots&\phi_{gn}(\g_n) \end{array}\right)\,.\ll{bnm}
\eq
is the so-called Brill-Noether matrix \cite{ne87}. Henceforth, we
shall restrict our attention to the special divisors $D_s$, such that
coefficients in the expansion (\ref{wd}) are independent on the point
$\g_j$
\[d\w_k=\phi_k(x_j)\,dx_j\,.\]
In this case all rows of the symmetric Brill-Noether matrix depend on
local coordinate $\{x_1,\ldots,x_n\}$ identically.

The Jacobi inversion problem (\ref{stinv})) is formulated as follows: for
a given point
\[\bu=(\b_1,\b_2\,,\ldots,\b_n)\in J(\C)\] find $n$ points
$\g_1,\g_2\,,\ldots,\g_n$ on the genus $g$ Riemann surface $\C$ such that
\bq
\sum_{k=1}^g\int_{\g_{0}}^{\g_k} d\w_j=\b_j\,,\quad j=1,\ldots,n.\ll{gjac}
\eq
Here we shall tacitly assume that the base point $\g_0\in\C$ has
already been fixed \cite{du81}.

If $n=g$ for almost all points $\bu\in J(\C)$ the solution
$D=\g_1+\cdots+\g_n$ exist and is uniquely determined by system
(\ref{gjac}) (for the unordered set of points $\g_j$) \cite{du81}.
However, if the degree $n<g$ of the symmetric product $\C^{(n)}$ is less
than genus $g$ of $\C$, the Abel map is lack of uniqueness. In this case
we can propose that two different points $\bu,\te{\bu}\in J(\C)$ have the
one Abel preimage $\{\g_1,\ldots,\g_n\}\in \C^{(n)}$.

The Abel preimage of the point $\bu\in J(\C)$ is given by set
$\{(p_1,q_1),\ldots,(p_n,q_n)\}\in \C^{(n)}$, where $\{q_1,\ldots,q_n\}$
are zeros of the Bolza equation \cite{bol95,bel97}
\bq
e(\l,\bu)= \l^n-\l^{n-1}\wp_{n, n}( \bu)-\l^{n-2}\wp_{n, n-1}(\bu)
-\ldots-\wp_{n, 1}(\bu)=0, \ll{coor}
\eq
and $\{p_1,\ldots,p_n\}$ are equal to
\bq
p_k=-\left.\dfrac{\pp\,e(\l,\bu)}{\pp\,\b_n}
\right|_{\,\l=q_k}.\ll{mom}
\eq
Here vector $\bu$ belongs to Jacobian $J(\C)$ and $\wp_{k,j}(\bu)$ is
the Kleinian $\wp$-function \cite{bol95,bel97}.

Now we turn to the uniform St\"{a}ckel systems. We can regard each
expression (\ref{stc}) as being defined on the genus $g$ Riemann
surface
\bq
\C:\quad
y_j^2=F(\l)\,,\qquad F(\l)=\sum_{k=1}^n \a_k\sm_{kj}(\l)-U(\l)\,,
\ll{sthc}
\eq
which depends on the values $\a_k$ of integrals of motion. For the
St\"{a}ckel systems on $\bR^{2n}$ the minimum admissible genus $g$ of the
curve $\C$ is equal to $g=[(n-1)/2]$.

The $n$th symmetric product of $\C$ defines the $n$-dimensional
Lagrangian submanifold in the complete symplectic manifold $\bR^{2n}$
\bq
\C^{(n)}:\qquad \C(p_1,q_1)\times\C(p_2,q_2)
\times\cdots\times\C(p_n,q_n)\,.\ll{stl}\eq
Then, the integration problem (\ref{stinv1}-\ref{stinv}) for equation
of motion is reduced to inverse Jacobi problem (\ref{rajm}) on
Lagrangian submanifold (\ref{stl}). The corresponding holomorphic
differentials $\d\w_k$ are equal to
\bq
d\w_k=\dfrac{\sm_{kj}(\l)\, d\l}{z(\l)}\,.\ll{stdif}
\eq
The set of these differentials either form a basis in the space of
holomorphic differentials $\H_1(\C)$ \cite{du81} or may be complement to
a basis. The corresponding $n\times n$ St\"{a}ckel matrix be the $n\times
n$ block of the transpose Brill-Noether matrix $\U_D^{*t}$.

The different blocks are determined the dual St\"{a}ckel systems. In this
case vectors differing the first entry only
\[\bu=\{t,\b_2,\ldots,\b_n\}\in J(\C)\,,\qquad
\te{\bu}=\{\te{t},\b_2,\ldots,\b_n\}\in J(\C)\]
have a common Abel preimage $\{(p_1,q_1),\ldots,(p_n,q_n)\}\in\C^{(n)}$.

Let us consider the standard basis of holomorphic differentials in
$\H_1(\C)$
\bq
d\w_j=\dfrac{\l^{j-1}}{z(\l)}\,d\l\,,\qquad j=1,\ldots,g\,.\ll{bd0}
\eq
Recall, that derivative $\U_D^*$ bears a great resemblance to the
canonical map $\C\to\bP^{g-1}$ and, therefore, to the Veronese map
$\bP^1\to\bP^{g-1}$ given by a basis for the polynomial ring of degree
$g-1$. With respect to the basis of $\H_1(\C)$ (\ref{bd0}), the Veronese
map of $\C$ has an extremely simple expression
\[(y,\l)\to
\l\to[\l^{g-1},\l^{g-2}\,\ldots,\l,1]\,.\]
By using the corresponding symmetric Brill-Noether matrix $\U_D^*$
(\ref{bnm}), we shall determine the St\"{a}ckel matrices as ($n\times n$)
blocks of the following ($g\times n$) matrix
\bq
\left(\begin{array}{cccc}
q_1^{g-1}&q_2^{g-1}&\cdots&q_n^{g-1}\\
\\
q_1^{g-2}&q_2^{g-2}&\cdots&q_n^{g-2}\\
\vdots&\ddots&\vdots&\vdots\\
1&1&\cdots&1\end{array}\right)\,.\ll{obnm}
\eq
Evidently, all the St\"{a}ckel matrices can not be obtained from the
symmetric Brill-Noether matrices. For instance, the St\"{a}ckel
matrices (\ref{sh}) do not belongs to the set of symmetric matrices.


\section{Lax representation.}
\setcounter{equation}{0}
Henceforth, we shall restrict our attention to the basis (\ref{bd0}) and
the symmetric matrix (\ref{obnm}). For the corresponding St\"{a}ckel
systems let us look for the Lax representation as
\bq
L=\left({\begin{array}{cc}h(\l,\p,\q)&
e(\l,\q)\\f(\l,\p,\q)&-h(\l,\p,\q)\end{array}}\right)\,.
\ll{lm}
\eq
Hereafter, by abuse of notation, we shall omit the some arguments at
the entries of the Lax matrix.

Let us fix hyperelliptic genus $g$ curve $\C$ and dimension of the phase
space $n\leq g$. Then we extract the ($n\times n$) St\"{a}ckel matrix
$\bs$ from the matrix (\ref{obnm}) and define the Hamilton function $H$
(\ref{ham}) with $U=0$.

To construct the Lax matrix let us determine function $e(\l,\bu)$
(\ref{coor}) initially
\bq
e(\l,\q)=\prod_{j=1}^n\,(\l-q_j)\,,\ll{el}
\eq
with $n$ zeroes, which are solution of the inverse Jacobi problem.

In the second step let us introduce the second entry of the Lax matrix as
\[h(\l)=-\dfrac1{2v(\l,\q)}\,\dfrac{d\,e(\l)}{d\,t}+w(\l,\p,\q)\,e(\l)\,.\]
Here function $v(\l,\q)$ is calculated by using the second Bolza equation
(\ref{mom})
\bq
\left.h(\l)\right|_{\,\l=q_k}=p_k=
\Bigl(\,\dfrac1{2v}\,\dfrac{d\,e(\l)}{d\,t}\,\Bigr)_{\,\l=q_k}=
-\left.\,\dfrac{\pp\,e(\l)}{\pp\,u_n}\right|_{\,\l=q_k}\,.\ll{sepv}\eq
Let the third entry of the Lax matrix takes the form
\[f(\l)=\dfrac{1}{v}\,\dfrac{d\,h(\l)}{dt}\,.\]
Here the single unknown function $w(\l,\p,\q)$ is determined such, that
the spectral curve of the Lax matrix (\ref{lm})
\bq
\C:\qquad z^2=F(\l)=-\det\,L_0(\l)=h^2(\l)+e(\l)\,f(\l)\ll{laxc}
\eq
be the same as initial algebraic curve $\C$ (\ref{stc}) by $U=0$.

The constructed above matrix $L_0(\l)$ (\ref{lm}) reads as
\bq
L_0(\l)=
\left({\begin{array}{ccc} -\dfrac{1}{2v}\,e_t(\l)+w(\l,\p,\q)\,e(\l)
&~&e(\l)\\
\\
\dfrac{1}{v}\,h_t(\l)&~&
\dfrac{1}{2v}\,e_t(\l)-w(\l,\p,\q)\,e(\l)\end{array}}\right)\,,\ll{lax0}
\eq
where
\[e_t=\dfrac{d\,e(\l)}{d\,t}=\{\,H,e(\l)\,\}\,,\qquad
h_t=\dfrac{d\,h(\l)}{d\,t}= \{\,H,h(\l)\,\}\,,\]
obeys the Lax equation
\[\dfrac{dL_0}{dt}=\{\,H,L_0\,\}=\Bigl[\,A_0,L_0\,\Bigr]\]
with the second matrix
\[A_0=v(\l,\q)\,\left({\begin{array}{cc}w(\l,\p,\q)&1
\\ 0&-w(\l,\p,\q)\end{array}}\right)\,.\]
By definition of the Lax matrix all the pairs of separation variables
$\g_j=(p_j,q_j)$ (\ref{el}-\ref{sepv}) lie on the spectral curve $\C$
(\ref{laxc}) of the matrix $L_0$ (\ref{lax0})
\[
z^2(\g_j)=p_j^2=\left.h^2(\l)\right|_{\l=q_j}=
F(\l=q_j)=\left.F(\l)\right|_{\g_j}\,.
\]
For the systems with polynomial potential $U\neq 0$ we propose to change
the entry $f(\l)$ in (\ref{lax0}) as
\[f(\l)=\dfrac{1}{v}\,\dfrac{d\,h(\l)}{dt}+u(\l,\q)e(\l)\,,\]
where we add new function $u(\l,\q)$ depending on coordinates only. Of
course, to construct the Lax matrix here
\bq
L(\l)=
\left({\begin{array}{ccc} -\dfrac{1}{2v}\,e_t(\l)+w(\l,\p,\q)\,e(\l)
&~&e(\l)\\
\\
\dfrac{1}{v}\,h_t(\l)+u(\l,\q)\,e(\l)&~&
\dfrac{1}{2v}\,e_t(\l)+w(\l,\p,\q)\,e(\l)\end{array}}\right)\,.\ll{lax}
\eq
we have to use the complete Hamiltonian with $U\neq 0$. The associated
second Lax matrix reads as
\bq
A=A_0+\left({\begin{array}{cc}0&0
\\ v(\l,q)\,u(\l,\q)&0\end{array}}\right)=
v(\l,\q)\,\left({\begin{array}{cc}w(\l,\p,\q)&1
\\ u(\l,\q)&-w(\l,\p,\q)\end{array}}\right)\,.
\ll{alax}
\eq
To consider the corresponding Lax equation
\[\dfrac{dL(\l)}{dt}=\Bigl[\,A(\l),L(\l)\,\Bigr]\,,\]
we can assume that the common factor $v(\l,\q)$ in front of the matrix
$A$ may be associated to the change of the time for the St\"{a}ckel
systems.

In general the proof of existence functions $v$, $w$ and $u$ requires
an application of the method of algebraic geometry \cite{bel97}. By
definition of the Lax matrices $L(\l)$ (\ref{lax}) and $A(\l)$
(\ref{alax}) this problem may be reduced to the solution of the single
equation
\bq
\dfrac{d\,f(\l)}{d\,t}-2v\,(u\,h-w\,f)=0\,,\quad
\Longleftrightarrow\quad
\dfrac{d F(\l,e,v,u)}{d\,t}=0\,,\ll{eq1}
\eq
for the given function $e(\l)$ (\ref{el}) and the given Hamiltonian
$H$ (\ref{ham}).

If we consider the lower ($n\times n$) block of the matrix (\ref{obnm}),
the differentials (\ref{stdif}) span a whole space $\H_1(\C)$ and the
Abel map is the one-to-one correspondence. In this case from equations
(\ref{mom}) and (\ref{sepv}) follows that
\[v_t(\l,q)=0\,,\qquad w(\l,\p,\q)=0\,.\]
If we put $v=1$, rename $t=x$  and introduce "new" time variable $\tau$,
the equation (\ref{eq1}) is rewritten as
\bq
\dfrac{\pp u(x,\tau,\l)}{\pp\tau}=\left[\dfrac14\partial^3_x +u(\l)\partial_x
+\dfrac12 u_{x}(\l)\,\right]\cdot e(\l)=0\,,\quad x=t\,,
\ll{eq2}
\eq
This equation may be identified with equation on the finite-band
stationary solutions $\frac{\pp u(x,\tau,\l)}{\pp\tau}=0$ of the
nonlinear soliton equations. In this theory equation (\ref{eq2}) is
called the generating equation. For different choices of the form of
$e(\l)$ and $u(\l)$, this procedure leads to different hierarchies of
integrable equations, as an example to the KdV, nonlinear Shr\"{o}dinger
and sine-Gordon hierarchies or to the Dym hierarchy (see references
within \cite{ts97d}).

Function $u(\l,\q)$ in (\ref{eq2}) is constructed by using function
$e(\l)$ (\ref{coor}-\ref{el})
\bq
u(\l,q_1,\ldots,q_n)=\Bigl[\phi(\l)e^{-2}(\l)\Bigr]_{MN}\,.\ll{u}
\eq
Here $\phi(\l)$ is a parametric function on spectral parameter and
$[\xi]_{N}$ is the linear combinations of the following Taylor
projections
\bq
{[ \xi ]_{N}}=\left[\sum_{k=-\infty}^{+\infty} z_k\l^k\,
\right]_{N}\equiv
\sum_{k=0}^{N} \xi_k\l^k\,,
\ll{cutmn}
\eq
or the Laurent projections \cite{ts96b,ts97d}.

If the differentials (\ref{stdif}) span the whole space $\H_1(\C)$ the
corresponding St\"{a}ckel systems describe all the possible systems,
which separable in the orthogonal curvilinear coordinate systems in
$\bR^n$ \cite{ts97d}. Let us consider the St\"{a}ckel systems which are
dual to these systems. To apply equation (\ref{tk}) to the function
$e(\l)$ (\ref{el}) and by using definition (\ref{sepv}) one gets
\ben
p_k&=&\left.\te{h}(\l)\right|_{\,\l=q_k}=
\left(\,-\dfrac{1}{2\te{v}}\,\{\te{H},e(\l)\}\,\right)_{\,\l=q_k}=
\dfrac{\det\bs}{\det\te{\bs}}\,
\left(\,-\dfrac{1}{2\te{v}}\,\{H,e(\l)\}\,\right)_{\,\l=q_k}\,\nn\\
\nn\\
\ll{v}\\
&=&
\dfrac{\det\bs}{\det\te{\bs}}\,
\left(\,\dfrac{v}{\te{v}}\,h(\l)\,\right)_{\,\l=q_k}=
p_k\,\dfrac{\det\bs}{\det\te{\bs}}\,\left(\dfrac{v}{\te{v}}\,
\right)_{\,\l=q_k}\,\nn
\en
Recall that $v=1$ for the integrable system with the Hamiltonian $H$
associated to the lower ($n\times n$) block of the matrix (\ref{obnm}).

Thus, according to (\ref{v}), below we shall consider the St\"{a}ckel
systems with   following functions $v(\q)$ only
\[v(q)=\dfrac{\det\bs(q_1,\ldots,q_n)}{\det\te{\bs}(q_1,\ldots,q_n)}.\]
The corresponding change of the time (\ref{tk}) depending on coordinates
only is closed to the Kolosoff transformation (\ref{tkow}) \cite{kol01}.

Let us briefly discuss canonical transformation which transforms a
Hamiltonian (\ref{ham}) to the natural form $H=T+V$. For integrable
systems separable in the orthogonal curvilinear coordinate systems on
$\bR^n$ the Abel map is one-to-one correspondence and $v_t=\{H,v\}=0$. In
this case we can put $v=1$ and introduce function $\B(\l)$
\bq
\B^2(\l)=e(\l)\,,\ll{bfl}
\eq
which was proposed in the theory of the soliton equations \cite{rw92}. It
allows us to rewrite generating function of integrals of motion
\bq
F(\l)=-{\B}^3\,{\B}_{tt}+u(\l,\q)\,{\B}^4\,.\ll{dsub}
\eq
as a Newton equation for the function $\B$
\bq
\ddot{\B}(\l,\q)=-F(\l,\a_1,\ldots,\a_n)\,{\B}^{-3}(\l,\q)
+u(\l,\q)\,{\B}(\l,\q)\,.\ll{newton}
\eq
To expand function $\B(\l)$ at the Laurent set
\[{\B}=\sum_{j=0}^N x_{j}\, \l^j\]
it is easy to prove that coefficients $x_j$ obey the Newton equation of
motion (\ref{newton}) (see references within \cite{rw92,ts97d}). Here we
reinterpret the coefficients of the function $F(\l)$ in (\ref{newton})
not as functions on the phase space, but rather as integration constants
$\a_j$ (\ref{alp}).

In general by $v_t\neq 0$ the generating function $F(\l)=-\det\,L(\l)$
(\ref{lax}) is equal to
\[F(\l)=\dfrac{1}{4v^2}\,\Bigl(e_t^2-2\,e\,e_{tt}\Bigr)
+\Bigl(\dfrac{v_t}{2v^2}-w\Bigr)\,\dfrac{e_t\,e}{v}
+\Bigl(w^2+\dfrac{u}{v}\Bigr)\,e^2\,.
\]
In this case the suitable canonical transformations, which transforms
any Hamiltonian (\ref{ham}) to the natural form, are unknown.

Although we can not proof validity of the presented Lax representation in
general, this construction works for the many well-known mechanical
systems. In the next Section we consider some two-dimensional St\"{a}ckel
systems in detail.


\section{Examples.}
\setcounter{equation}{0}
Let us consider four orthogonal systems of coordinates on plane:
elliptic, parabolic, polar and cartesian \cite{gol80}. The Lax matrix
$L_0(\l)$ (\ref{lax0}) by $U=0$ is transformed to the Lax matrix $L(\l)$
(\ref{lax}) by $U\neq 0$ by using the outer automorphism of the space of
infinite-dimensional representations of underlying algebra $sl(2)$
\cite{ts96b,ts97d}. Since, we shall consider the Lax representations for
the geodesic motion by $U=0$ more extensively.
\vskip0.4cm
\par\noindent
{\it 1. Parabolic and cartesian coordinate systems ($w(\l,\p,\q)=0$).}
\vskip0.4cm
\par\noindent
Let us consider two hyperelliptic curves
\ben
&\C^{(1)}:\qquad &z^2=\prod_{i=1}^{2g+1} (\l-\l_i)\,,\nn\\
\ll{c}\\
&\C^{(2)}2:\qquad &z^2=\l^{-1}\,{\prod_{i=1}^{2g+1} (\l-\l_i)}\,.\nn
\en
If we choose the standard basis in the space of holomorphic
differentials one gets the following symmetric matrices (\ref{obnm})
for two-dimensional systems
\bq
\U_1^{*t}(q_1,q_2)=
\left(\begin{array}{cc}
q_1^{g-1}&q_2^{g-1}\\
\vdots&\vdots\\
q_1^2&q_2^2\\
\\q_1&q_2\\
\\-1&-1\end{array}\right)\,,
\qquad
\U_2^{*t}(q_1,q_2)=
\left(\begin{array}{cc}
q_1^{g-2}&q_2^{g-2}\\
\vdots&\vdots\\
q_1&q_2\\
\\ 1&1\\
\\-\dfrac{1}{q_1}&-\dfrac{1}{q_2}\end{array}\right)
\ll{bnmp}
\eq
Different ($2\times 2$) blocks of the matrices $\U_j^{*t}$ determine
different St\"{a}ckel systems.

Let us consider two blocks for the each matrices, such that the
corresponding change of the time will be same as the Kolosoff
transformation (\ref{tkow}) \cite{kol01}. So, for the curve $\C^{(1)}$
we shall consider the following matrices
\bq
\bs_1=\left({\begin{array}{cc}q_1& q_2\\-1&-1\end{array}}\right)\,,\qquad
\te{\bs}_1=\left({\begin{array}{cc}q_1^2&q_2^2\\-1&-1\end{array}}\right)\,.
\ll{st1}
\eq
For the second curve $\C^{(2)}$ the associated St\"{a}ckel matrices
are equal to
\bq
\bs_2=\left({\begin{array}{cc}1&1\\ \\-\dfrac{1}{q_1}&-\dfrac{1}{q_2}
\end{array}}\right)\,,\qquad
\te{\bs}_2=\left({\begin{array}{cc}q_1&q_2\\ \\-\dfrac{1}{q_1}&-\dfrac{1}{q_2}
\end{array}}\right)\,.
\ll{st2}
\eq
Introduce the Hamilton functions (\ref{ham}) by $U=0$
\ben
H_0^{(1)}&=&\dfrac{p_1^2 -p_2^2}{q_1-q_2}\,,\qquad
\te{H}_0^{(1)}=(q_1+q_2)^{-1}\,H_0^{(1)}\,,\nn\\
\ll{ham12}\\
H_0^{(2)}&=&\dfrac{q_1\,p_1^2 -q_2\,p_2^2}{q_1-q_2}\,,\qquad
\te{H}_0^{(2)}=(q_1+q_2)^{-1}\,H_0^{(2)}\,.\nn
\en
The corresponding second integrals of motion of the dual systems are
related
\[
\te{J}_0^{(k)}=J_0^{(k)}-\dfrac{q_1\,q_2}{q_1+q_2}\,H_0^{(k)}\,,
\qquad k=1,2\,.\]
The functions $e(\l,\bu)$ (\ref{coor})
\bq
e_1(\l)=(\l-q_1)(\l-q_2)\,,\qquad
e_2(\l)=\dfrac{(\l-q_1)(\l-q_2)}{\l}\,.\ll{el12}
\eq
have two zeroes, which are solution of the inverse Jacobi problem
(\ref{stinv}) on $\C^{(1)}$ and $\C^{(2)}$, respectively.

Let us introduce the new  physical variables at once. For the first curve
$\C^{(1)}$ equation (\ref{newton})
\[e_1(\l)=(\l-q_1)(\l-q_2)=\B^2(\l)\,,\qquad\B(\l)=
\l-\dfrac{x}{2}-\dfrac{y}{4\l}\,,\]
immediately yields the following canonical transformation
\ben
q_1&=&\dfrac{x-\sqrt{2\,y}}2\,,\qquad p_1=p_x-\sqrt{2\,y}\,p_y\,,\nn\\
q_2&=&\dfrac{x+\sqrt{2\,y}}2\,,\qquad p_2=p_x+\sqrt{2\,y}\,p_y\,.\nn
\en
These variables obviously related to the cartesian coordinate system.
For the second curve $\C^{(2)}$ the corresponding equation
\[e_2(\l)=\l^{-1}\,(\l-q_1)(\l-q_2)=\l-x-\dfrac{y^2}{4\l}\]
defines the standard parabolic coordinate system
\ben
q_1&=&\dfrac{x-\sqrt{x^2+y^2}}2\,,\qquad
p_1=p_x-\dfrac{\sqrt{x^2+y^2}+x}{y}\,p_y\,,\nn\\
q_2&=&\dfrac{x+\sqrt{x^2+y^2}}2\,,\qquad
p_2=p_x+\dfrac{\sqrt{x^2+y^2}-x}{y}\,p_y\,.\nn
\en
By $U=0$ the Hamilton functions are given by
\[H_0^{(1)}=4\,p_x\,p_y\,,\qquad H_0^{(2)}=p_x^2+p_y^2\,.\]
According to (\ref{sepv}) and (\ref{v}) functions $v(q_1,q_2)$
entering in the Lax representation are equal to
\ben
v(\l,q_1,q_2)&=&1\,\quad \mbox{\rm for matrices}\quad
\bs_{1,2}\,,\nn\\
\ll{v12}\\
v(\l,q_1,q_2)&=&(q_1+q_2)^{-1}=\dfrac{1}{x}\,\quad
\mbox{\rm for matrices}\quad
\te{\bs}_{1,2}\,.\nn
\en
In physical variables the Lax matrices are given by
\ben
L_0^{(1)}&=&\left({\begin{array}{cc} p_x+(2\l-x)\,p_y&
\l^2-\l\,x+\dfrac{x^2-2\,y}{4}\\ \\
-4\,p_y^2&-p_x-(2\l-x)\,p_y\end{array}}\right)\,,\nn\\
\ll{lax12}\\
L_0^{(2)}&=&\left({\begin{array}{cc} p_x+\dfrac{1}{2\l}\,y\,p_y&
\l-x-\dfrac1{4\l}\,y^2\\ \\
\dfrac1{\l}\,p_y^2&-p_x-\dfrac{1}{2\l}\,y\,p_y\end{array}}\right)\,.\nn
\en
For the dual St\"{a}ckel systems the Lax matrices $\te{L}_0^{(1,2)}$
have the form
\ben
\te{L}_0^{(1)}&=&L_0^{(1)}+\left({\begin{array}{cc} 0&0\\ \\
4\dfrac{p_x\,p_y}{x}&0\end{array}}\right)=
L_0^{(1)}+\left({\begin{array}{cc} 0&0\\ \\
\te{H}_0^{(1)}&0\end{array}}\right)
\,,\nn\\
\ll{dlax12}\\
\te{L}_0^{(2)}&=&L_0^{(2)}+\left({\begin{array}{cc} 0&0\\ \\
\dfrac{p_x^2+p_y^2}{x}&0\end{array}}\right)=
L_0^{(1)}+\left({\begin{array}{cc} 0&0\\ \\
\te{H}_0^{(2)}&0\end{array}}\right)\,.\nn
\en
By using property $\{h_t(\l),v(\q)\}=0$ of the function $v(\q)$
(\ref{v12}) we can easy proof equation (\ref{eq1}) for the dual systems
by using the same equation for the system with $v_t=0$
\[\te{f}_t=\{\te{H},f+\te{H}\}=f_t=0\,.\]
Another consequence of this property is that the function $w(\l,\p,\q)$
in (\ref{lax0}) is equal to zero.

Note, in the works \cite{avm88} and \cite{hh87}, devoted to the
Kowalewski top, the common Lax matrices are proposed for the both dual
systems after the non-canonical change of variables. Here we obtain
different Lax matrices for the systems connected by non-canonical change
of the time.

The spectral curves of the matrices $L_0$ (\ref{lax12}) coincides with
the initial curves $\C_0^{(1,2)}$ (\ref{stc}) by $U=0$
\bq
z^2=H_0^{(1)}\,\l+J_0^{(1)}\,,\qquad
z^2=H_0^{(2)}+\dfrac{J_0^{(2)}}{\l}\,.\ll{c12}
\eq
Here $J_0^{(1,2)}$ be the second integrals of motion (\ref{int}). For
the dual systems with the Hamilton functions $\te{H}_0^{(1,2)}$ the
corresponding spectral curves are equal to
\bq
z^2=\te{H}_0^{(1)}\,\l^2+\te{J}_0^{(1)}\,,\qquad
z^2=\te{H}_0^{(2)}\l+\dfrac{\te{J}_0^{(2)}}{\l}\,.\ll{dc12}
\eq
If for the system with the Hamiltonian $H_0^{(1)}$ the Abel map is
one-to-one correspondence on the curve (\ref{c12}), then for the same
system on the curve
\[z^2=e_2\,\l^2+e_1\,\l+e_0\]
the associated Abel map is lack of uniqueness in general. So, on this
curve we can introduce the second St\"{a}ckel system with the dual
Hamiltonian $\te{H}_0^{(1)}$.

Let us briefly consider systems with polynomial potentials $U\neq )$. As
an example, introduce different potentials for the curves $\C^{(1,2)}$
(\ref{c})
\bq
U^{(1)}(q_j)=\a^2\,q_j^5+\b\,q_j^3\,,\qquad
U^{(1)}(q_j)=\a^2\,q_j^3+\b\,q_j\,.\ll{pot12}
\eq
To describe these potentials we have to put $N=6$ and $N=4$ in
(\ref{cutmn}) and have to use the following parametric functions
\[\phi^{(1)}(\l)=-\a^2\,\l^5\quad\mbox{\rm and}\quad\phi^{(2)}(\l)=-\a^2\,\l^3\]
for the curves $\C^{(1)}$ and $\C^{(2)}$, respectively. For the both
curves the common function $u(\l,q_1,q_2)$ is given by
\bq
u^{(1,2)}=-\a^2\,(\l+2\,x)\,.\ll{u12}
\eq
Here we restrict ourselves the presentation of the function $u$ only, the
complete Lax matrices $L(\l)$ may be constructed by the rule (\ref{lax}).

The spectral curves of the corresponding matrices (\ref{lax}) coincides
with the initial curves (\ref{sthc}). For instance, curves for the
systems with dual Hamiltonians $\te{H}^{(1,2)}$ are
\ben
&&\C^{(1)}:\qquad z^2=\a^2\,\l^5+\b\,\l^3-\te{H}\,\l-\te{J}\,,\nn\\
\ll{pc12}
&&\C^{(2)}:\qquad z^2=\a^2\,\l^3-\te{H}\,\l+\b-\dfrac{\te{J}}{\l}\,.\nn
\en

The Poisson bracket relations for the Lax matrix
(\ref{lax12}-\ref{dlax12}) are closed into the following linear
$r$-matrix algebra
\ben
&&\{{\on{L}{1}}(\l),{\on{L}{2}}(\m)\}=
[r_{12}(\l,\m),{\on{L}{1}}(\l)]-[r_{21}(\l,\m),{\on{L}{2}}(\m)\,]\,,\nn\\
\ll{rpoi}\\
&&r_{21}(\l,\mu)=-\Pi\,r_{21}(\l,\mu)\,\Pi\,.\nn
\en
Here the standard notations are introduced:
\[{\on{L}{1}}(\l)= L(\l)\otimes I\,,\qquad
{\on{L}{2}}(\m)=I\otimes L(\m)\,,\] and $\Pi$ is the permutation
operator of auxiliary spaces \cite{ft87}.

By $v_t=0$ for the systems related to the  matrices  $\bs^{1,2}$ the
corresponding $r$-matrices $r_{ij}(\l,\m)$ in (\ref{rpoi}) consist of two
terms
\bq
r_{ij}=r^{p}_{ij}+r^{u}_{ij}\,.\ll{r2}
\eq
The first matrix is a standard $r$-matrix on the loop algebra
$\L(sl(2))$
\bq
r^{p}_{12}(\l,\m)=\dfrac{\Pi}{\l-\m}=
\dfrac1{\l-\m}\,
\left({\begin{array}{cccc} 1&0&0&0\\0&0&1&0\\ 0&1&0&0\\0&0&0&1
\end{array}}\right)
\,.\ll{r}
\eq
The second matrix may be associated to outer automorphism of the space
of infinite-dimensional representations of underlying algebra $sl(2)$
\cite{ts96b,ts97d}. The corresponding dynamical $r^{u}_{ij}$-matrices
depend on the coordinates only
\bq
 r^{u}_{12}=
\dfrac{u(\l,\q)-u(\m,\q)}{\l-\m}\,\s_-\otimes\s_-\,,
\qquad
\s_-=\left(\begin{array}{cc} 0&0\\1&0\end{array}\right)\,.\nn
\eq

By $v_t\neq 0$ for the dual St\"{a}ckel systems related to the matrices
$\te{\bs}_{1,2}$ we have to add to the $r$-matrices $\ref{r2}$ the third
term
\bq
r^{v}_{12}=v(q_1,q_2)\,\left({\begin{array}{cccc} 0&0&0&0\\0&0&0&0\\
0&1&0&0\\ 0&0&0&0\end{array}}\right)\,, \ll{rv}
\eq
where the second matrix $r^{v}_{21}$ is defined by (\ref{rpoi}).

The matrix $r^{v}_{ij}$ may be connected with the Drinfeld twist for the
Toda lattice associated to the root system $\D_n$. Let us consider the
Drinfeld twist \cite{kst96} of the quantum $R$-matrix
\bq
\te{R}=F\,R\,F_{21}^{-1}\,,\qquad F_{21}=\Pi\,F\,\Pi\,.
\ll{tR}
\eq
Here matrix $R$ satisfies the Yang-Baxter equation and matrix $F$ has the
special property \cite{kst96}. To introduce the corresponding linear
$r$-matrix \cite{ts94}, one gets
\[R=I+2\eta\,r^p+O(\eta^2)\,,\qquad F=I+\eta\,r^v+O(\eta^2)\,.\]
Then we consider limit of the twisted matrix $\te{R}$ by $\eta\to 0$
\bq
\te{R}_{12}=I+\eta\,\Bigl(\,r^p_{12}+r^v_{12} -
\Pi\,(\,r^p_{12}+r^v_{12}\,)\,\Pi\,\Bigr)+ O(\eta^2)\,.
\ll{twr}
\eq
Formally, coefficients by $\eta$ may be called twisted linear
$r$-matrix.

By using generators $\h,\e,\f$ of the underlying Lie algebra $sl(2)$
\bq [\h,\e]=2\e\,,\qquad [\h,\f]=-2\f\,,\qquad [\e,\f]=\h\,,\ll{slt}
\eq
let us introduce an appropriate element $\F\in U(sl(2))\otimes
U(sl(2))$
\[\F_\xi=exp(\xi\cdot\e\otimes\f)\,,\qquad \xi\in\bC\,\]
belonging to a tensor product of the corresponding universal
enveloping algebras $U(sl(2))$ \cite{kst96}. In the fundamental
spin-1/2 representation $\rho_{1\over2}$ we have
\[F(\xi) =(\rho_{1\over2}\otimes \rho_{1\over2})\F_\xi=
\left(\begin{array}{cccc}1&0&0&0\\ 0&1&0&0\\
0&\xi&1&0\\0&0&0&1\end{array}\right)\,.
\]
To substitute in (\ref{tR}) the Yang solution of the Yang-Baxter equation
$R=I+\dfrac{\eta}{\l}\,\Pi$ we get a twisted $R$-matrix. If the element
$\xi(\q)$ be a suitable function on coordinates,  this dynamical twisted
$R$-matrix may be used to description of the Toda lattice associated with
the $\D_n$ root system \cite{ts98a}.

Let us consider twisted dynamical matrix (\ref{tR}) by $\xi=v(\q)$. We
can see that the linear $r$-matrix associated to the dual St\"{a}ckel
system (\ref{r}-\ref{rv})
\[r_{12}=r^p_{12}+r^v_{12}\,,\qquad r_{21}=-\Pi\,(r^p_{12}+r^v_{12})\,\Pi\,\]
is equal to the half of the twisted linear matrix (\ref{twr}).

Recall, for the St\"{a}ckel matrices $\bs_1$ (\ref{st1}) and $\bs_2$
(\ref{st2}) the corresponding differentials (\ref{stdif}) span $\H_1$.
Since, the associated Hamilton functions have a natural form in physical
variables. For instance, Hamiltonians with potentials (\ref{pot12}) are
given by
\ben
H^{(1)}&=& 2\,p_x\,p_y + \dfrac{\a^2}{4}\,(y^2 +5\,x^{2}\,y +
\dfrac54\,x^{4}) +
\dfrac{\b}{2}\,(\dfrac32\,x^{2}+y)\,,\nn\\
\ll{nham1}\\
H^{(2)}&=&\dfrac{p_x^2}2+\dfrac{p_y^2}2+\dfrac{\a^2}2\,x\,(2x^2+y^2)+\b\,.\nn
\en
To consider the dual St\"{a}ckel systems we have to use additional
transformation
\bq
x=\sqrt{2\,\te{x}}\,,\qquad p_x=\te{p}_x\,\sqrt{2\,\te{x}}\,,\ll{ctr1}
\eq
for the first curve and the following more complicated transformation
\ben
x&=&\dfrac32\,\te{x}^{2/3}\,,\qquad p_x=\te{p}_x\,\te{x}^{1/3}\,,\nn\\
\ll{ctr2}
y&=&\sqrt{\dfrac23}\,\dfrac{\te{p}_y}{\a}\,,\qquad
p_y=-\sqrt{\dfrac32}\,\a\,\te{y}\,\nn
\en
for the second curve. After this canonical change of variables the
Hamiltonians $\te{H}^{(1,2)}$ (\ref{ham12}) obtain the natural form
\ben
\te{H}^{(1)}&=&  2\,\te{p}_x\, p_y
+\dfrac{\a^2\,(y^2+10\,y\,\te{x}+5\,\te{x}^{2})}{8}\,\sqrt{\dfrac2{\te{x}}}
+\dfrac{\b\,(3\,\te{x}+y)}{4}\,\sqrt{\dfrac2{\te{x}}}\,,\nn\\
\ll{nham2}\\
\te{H}^{(2)}&=&
\dfrac12 \,(\te{p}_x^2 + \te{p}_y^2) +
\dfrac {3\,\a^2}{4}\,\te{x}^{-2/3}\,(\dfrac92\,\te{x}^{2}+\te{y}^2)+\b\,\te{x}^{-2/3}
\,.\nn
\en
The system with the Hamiltonian $H^{(2)}$ is so-called second integrable
case of the Henon-Heiles system \cite{rgb89}. The dual system with the
Hamiltonian $\te{H}^{(2)}$ is so-called Holt-type system \cite{rgb89}.
Note, the second integral of motion is a polynomial of the fours order in
momenta for the Holt system.

Additional canonical transformation (\ref{ctr2}) allows us to get natural
Hamiltonians for the restricted class of the potentials $U$ (\ref{pot12})
only. Unlike canonical transformation (\ref{ctr1}) may be used for any
potentials $U$. As an example, rational potential
\[U(q)=\dfrac{\a}{q^2}+\dfrac{\b}{q}+
\g\,q+\d\,q^2+\rho\,q^4
\]
give rise the following Hamiltonian
\[
\te{H}=2\,\te{p}_x\,p_y
-\dfrac{4\,\a}{(\te{x}-y)^2}
-\dfrac{\b}{\te{x}-y}\,\sqrt{\dfrac2{\te{x}}}
+\dfrac{\g}4\,\sqrt{\dfrac2{\te{x}}}
 +\dfrac12\,\d+\dfrac{\rho}{2}\,(\te{x}+y)
\,.
\]
Also we can add potential terms (\ref{nham2}) to this Hamiltonian.

By $v=1$ and $w=0$ the Lax representation (\ref{lax0}) for a system with
an arbitrary number $n$ of degrees of freedom may be regarded as a
generic point at the loop algebra $\L(sl(2))$ in fundamental
representation after an appropriate completion \cite{ts97d}. As an
example, for the generalized parabolic coordinate systems function
$e(\l)$ is given by
\[
e(\l)=
\dfrac{\prod_{j=1}^n (\l-q_j)}{\prod_{k=1}^{n-1}(\l-\d_k)}=
\l-x_n+\sum_{k=1}^{n-1}\dfrac{x_k^2}{4\,(\l-\d_k)}\,,\qquad\d_k\in\bR\,.
\]
To construct the Lax representation for a potential motion we can use the
outer automorphism of the space of infinite-dimensional representations
of $sl(2)$ proposed in \cite{ts96b}.

By $v_t\neq 0$ for the dual St\"{a}ckel systems the Lax representations
may be constructed without any problem as well. For instance, let us
consider system with the three degrees of freedom. To construct the Lax
matrix  by (\ref{lax0}-\ref{lax}) with the function $u$ given by
(\ref{u12}) one gets
\ben
e(\l)&=&\l-x-\dfrac{y^2}{\l}-\dfrac{z^2}{4\,(\l-k)}\,,\qquad
k\in\bR\,,\nn\\
\nn\\
\te{H}&=&\dfrac1{x}\,\Bigl(\,p_x^2+p_y^2+p_z^2+\dfrac{a^2\,k\,z^2}4\,\Bigr)
+\dfrac{a^2}2\,\Bigl(\,2\,x^2+y^2+z^2\,\Bigr)\,.\nn
\en
After an additional canonical transformation (\ref{ctr2}) extended on the
$p_z,z$ variables the Hamilton function takes the form
\[
\te{H}=
\te{p}_x^2 + \te{p}_y^2+\te{p}_z^2\,\Bigl(1+\dfrac{k}3\,\te{x}^{-2/3}\,\Bigr) +
\dfrac{3\,\a}8\,\te{x}^{-2/3}\,(\dfrac92\,\te{x}^{2}+\te{y}^2+\te{z}^2\,)\,.
\]
So, the main  unsolved problem is to introduce additional canonical
transformation, which transform the dual Hamilton function $\te{H}$ into
the natural form.

\vskip0.4cm
\par\noindent
{\it 2. Elliptic and polar coordinates ($w(\p,\q)\neq 0$).}
\vskip0.4cm
\par\noindent
Recall, that the polar coordinate system may be obtained from elliptic
coordinate system and, therefore, we shall consider elliptic
coordinate systems in detail.

For the elliptic coordinate systems algebraic curve is given by
\[\C^{(3)}\qquad z^2=\dfrac{\prod_{i=1}^{2g+1} (\l-\l_i)}{(\l-k)(\l+k)}\,,\qquad
k\in\bC\,.\nn
\]
Let us consider two St\"{a}ckel matrices associated to this curve
\bq
\bs_3=\left({\begin{array}{cc}\dfrac{q_1}{q_1^2-k^2}&
\dfrac{q_2}{q_2^2-k^2}\\
\\ \dfrac{1}{q_1^2-k^2}& \dfrac{1}{q_2^2-k^2}
\end{array}}\right)\,,\qquad
\te{\bs}_3=\left({\begin{array}{cc}
\dfrac{4\,q_1^2}{q_1^2-k^2}& \dfrac{4\,q_2^2}{q_2^2-k^2}\\
\\ \dfrac{1}{q_1^2-k^2} & \dfrac{1}{q_2^2-k^2}
\end{array}}\right)\,,
\ll{st3}
\eq
The corresponding non-canonical change of the time (\ref{dham}) is closed
to the Kolosoff transformation (\ref{tkow}) \cite{kol01}.

For the polar coordinate system the Sta\"{a}ckel matrices are
non-symmetric matrices
\bq
\bs_4=\left({\begin{array}{cc}1&0\\
\dfrac1{q_1^2}& \dfrac{1}{4\,(q_2^2-k)}
\end{array}}\right)\,,\qquad
\te{\bs}_4=\left({\begin{array}{cc}
q_1^2& 0\\
\\ \dfrac{1}{q_1^2} & \dfrac{1}{4\,(q_2^2-k)}
\end{array}}\right)\,,
\ll{st4}
\eq
The corresponding non-canonical change of the time (\ref{dham}) is closed
to the Kepler change of the time \cite{gol80}.

By $U=0$ the initial hyperelliptic curves (\ref{sthc}) for the matrices
$\bs_3$ and $\te{\bs}_3$ are given by
\bq
z^2=\dfrac{H_0\,\l+J}{\l-k^2}\,,\qquad
z^2=\dfrac{4\,\te{H}_0\,\l^2+\te{J}_0}{\l-k^2}\,,\ll{c3}
\eq
with the following Hamiltonians
\bq
H_0^{(3)}=\dfrac{p_1^2\,(q_1^2-k^2) -
p_2^2\,(q_2^2-k^2)}{q_1-q_2}\,,\qquad
\te{H}_0^{(3)}=\dfrac1{4\,(q_1+q_2)}\,H\,,\ll{h3}
\eq
The Hamiltonians related to the matrices $\bs_4$ and $\te{\bs}_4$ read as
\bq
H_0^{(4)}=p_1^2-4\,\dfrac{q_2^2-k}{q_1^2}\,p_2^2\,,\qquad
\te{H}_0^{(4)}=q_1^{-2}\,H^{(4)}_0\,.
\ll{h4}
\eq
Let us fix elliptic coordinates by using equation
\[e(\l)=\dfrac{(\l-q_1)\,(\l-q_2)}{(\l-k)\,(\l+k)}
=1-\dfrac{x^2}{4(\l-k)}-\dfrac{y^2}{4(\l+k)}\]
such that
\ben
q_1&=&\dfrac{x^2+y^2}8 +
\dfrac12\,\sqrt{(x^2+y^2)^2+16\,k\,(x^2-y^2)+64\,k^2}\,,\nn\\
q_2&=&\dfrac{x^2+y^2}8
- \dfrac12\,\sqrt{(x^2+y^2)^2+16\,k\,(x^2-y^2)+64\,k^2}\,.\nn
\en
The corresponding equation for the polar coordinates
\[e(\l)=\dfrac{q_1\,(\l-q_2)}{\l\,(\l-1)}
=\dfrac{x^2}{4\,\l}+\dfrac{4\,y^2}{\l-1}\]
immediately yields
\[q_1=r=\sqrt{x^2+y^2\,}\,,\qquad q_2=\cos^2(\phi)=\dfrac{x^2}{x^2+y^2}\,.\]
In physical variables the Hamiltonians (\ref{h3}-\ref{h4}) have a
common form
\[
H={p_x^2+p_y^2}\,,\qquad
\te{H}=\dfrac{p_x^2+p_y^2}{x^2+y^2}\,.\]

To construct the Lax representations we begin with the calculations of
the functions $v(\l,\q)$ by the rule (\ref{sepv}- \ref{v})
\ben
v&=&1\qquad{\mbox{\rm for matrices}}\quad \bs_{3,4}\nn\\
\nn\\
v&=&\dfrac14\,(q_1+q_2)^{-1}=\dfrac1{x^2+y^2}
\qquad{\mbox{\rm for matrix}}\quad \te{\bs}_{3,4}\ll{v34}\\
\nn\\
v&=&\dfrac1{q_1^2}=\dfrac1{x^2+y^2}
\qquad{\mbox{\rm for matrix}}\quad \te{\bs}_{4}\,.\nn
\en
So, for the St\"{a}ckel systems associated with the matrices $\bs_3$
(\ref{st3}) and $\bs_4$ (\ref{st4}) one gets
\bq
L_0(\l)=\left({\begin{array}{ccc}
\dfrac{x\,p_x}{2\,(\l-k)}+\dfrac{y\,p_y}{2\,(\l+k)}&~&
\epsilon\l-\dfrac{x^2}{4\,(\l-k)}-\dfrac{y^2}{4\,(\l+k)}\\ \\
\dfrac{p_x^2}{\l-k}+\dfrac{p_y^2}{\l+k}&~&
-\dfrac{x\,p_x}{2(\l-k)}-\dfrac{y\,p_y}{2(\l+k)}\end{array}}\right)\,.
\ll{lax3}
\eq
Here $\epsilon=1$ for the elliptic coordinate system and $\epsilon=0$
for the parabolic coordinate system. The spectral curve of the Lax
matrix $L_0(\l)$ coincides to the initial curve (\ref{c3}).

For the dual system, in contrast to the cartesian and parabolic
coordinates, the Lax matrix has the more complicated form. Both these
Lax matrices may be constructed by the rule (\ref{lax}) with the
following common function $w(\p,\q)$
\bq
w(\p,\q)=2\,\sqrt{\te{H}\,}\,.\ll{w3}
\eq
The Lax matrix reads as
\bq
\te{L}_0(\l)=L_0(\l)+\left({\begin{array}{cc}
w\,e(\l)&0\\ \\ -2\,w\,\Bigl[\,h(\l)-w\,e(\l)-\epsilon\,w\,\Bigr]&
-w\,e(\l)\end{array}}\right)\,, \quad \epsilon=0,1\,.\ll{dlax3}
\eq
Here $e(\l)$ and $h(\l)$ are entries of the corresponding matrices
$L_0(\l)$ (\ref{lax3}) by $\epsilon=0,1$. As above, the spectral curve of
the Lax matrix $\te{L}_0(\l)$ by $\epsilon=1$ coincides with the initial
curve (\ref{c3}).

For the cartesian and parabolic coordinate systems we can get equation
\[
\Bigl\{\,\Bigl\{\,H,v^{-1}(\q)\,\Bigr\}\,,\,e(\l,\q)\,\Bigr\}=
\Bigl\{\,\Bigl\{\,H,(q_1+q_2)\,\Bigr\}\,,\,e(\l,\q)\,\Bigr\}=2\,,\]
on the Hamiltonian $H$, function $e(\l)$ and function $v(\q)$ defining
change of the time. For the polar and elliptic coordinate systems the
corresponding equation is
\[\Bigl\{\,\Bigl\{\,H,v^{-1}(\q)\,\Bigr\}\,,\,e(\l,\q)\,\Bigr\}=8\,
\Bigl(\,e(\l)-\epsilon\,\Bigr)\,,
\qquad \epsilon=0,1\,.\]
Hence, from the equation (\ref{eq1}) follows that the function $w(\p,\q)$
in (\ref{lax0}-\ref{w3}) does not equal to zero. If we consider more
complicated change of the time for the cartesian and parabolic coordinate
systems, one gets non-zero function $w$ (\ref{lax0}) as well.

The Poisson bracket relations for the Lax matrix $L_0(\l)$ (\ref{lax3})
are closed into the standard linear $r$-matrix algebra (\ref{rpoi}) with
rational $r$-matrix (\ref{r}) on the loop algebra $\L(sl(2))$
\cite{ft87}.

The Poisson brackets relations for the dual Lax matrix $\te{L}_0(\l)$
(\ref{dlax3}) have a poly-linear form
\ben
\{\,{\on{L}{1}}(\l)\,,\,{\on{L}{2}}(\m)\,\}&=&
\Bigl[\,r_{12}\,,\,{\on{L}{1}}(\l)\,\Bigr]+
\Bigl[\,r_{21}\,,\,{\on{L}{2}}(\m)\,\Bigr]\nn\\
\nn\\
&+&
R\,\,{\on{L}{1}}(\l)\,\,{\on{L}{2}}(\m)+{\on{L}{1}}(\l)\,\,{\on{L}{2}}(\m)\,R-
{\on{L}{1}}(\l)\,R\,\,{\on{L}{2}}(\m)-{\on{L}{2}}(\m)\,R\,\,{\on{L}{1}}(\l)\,.\nn
\en
Here linear $r$-matrix reads as
\[
r_{12}(\l,\m)=r^p_{12}(\l-\m)+4\,\epsilon\,r^w_{12}\,,\qquad
r_{21}(\l,\m)=-\Pi\,r_{21}(\l,\m)\,\Pi\,,
\]
where $r^p(\l-\m)$ be the standard linear $r$-matrix on the loop algebra
$\L(sl(2))$. The second dynamical term  is given by
\[
r^w_{12}=v(\q)\,\left({\begin{array}{cccc} 0&0&0&0\\0&0&0&0\\ w&1&0&0\\
0&-w&0&0\end{array}}\right)\,.\] The quadratic  $R$-matrix is closed to
the twisted linear $r$-matrix (\ref{twr})
\[
R=-\dfrac{2}{w}\,\Bigl(\,r^w_{12}+r^w_{21}\,\Bigr)=
-\dfrac{2}{w}\,\Bigl(\,r^w_{12}-\Pi\,r^w_{12}\,\Pi\,\Bigr)\,.\]

For the systems with $U\neq 0$ functions $u(\l,\q)$ may be constructed
as usual \cite{ts96b,ts97d}. Note, the both dual Hamiltonians obtain a
natural form after the following additional canonical transformation
of variables
\ben
x&=&\sqrt{\te{x}}-\sqrt{\te{y}}\,,\qquad
p_x=\sqrt{\te{x}}\,\te{p}_x-\sqrt{\te{y}}\,\te{p}_y\,,\nn\\
y&=&-i\,(\sqrt{\te{x}}+\sqrt{\te{y}})\,,\qquad
p_y=i\,(\sqrt{\te{x}}\,\te{p}_x+\sqrt{\te{y}}\,\te{p}_y)\,.\nn
\en
As an example, for elliptic coordinate system the uniform potential
\[U^{(3)}(q_j)=\a\,q_j^2+\b\,q_j\,\]
give rise to the following dual Hamiltonian
\[\te{H}=2\,{\te{p}_x\,\te{p}_y}+
\dfrac\a4\,(\te{x}+k)\,(\te{y}+k)-
\dfrac{\b}{8\,\sqrt{\te{x}\,\te{y}\,}}\,\,
(2\,\te{x}\te{y}+k\,\te{x}+k\,\te{y})\,.\] For the polar coordinate
system we present the non-uniform degenerate potentials
\[U^{(4)}_1(q_1)=\b\,,\qquad U^{(4)}_2(q_2)=0\,,\]
associated to the dual Hamiltonians in the form
\[
\te{H}=\dfrac{\te{p}_x\,\te{p}_y}2-\b\,(\,
\dfrac{16}{\sqrt{\te{x}\,\te{y}\,}}+
\dfrac1{\te{x}}+\dfrac{1}{\te{y}}-\dfrac{2}{\te{x}\,\te{y}}\,)\,.\]
Both these systems may be considered as an integrable deformation of
the Kepler problem.

\section{Conclusions}
In this paper we have considered the non-canonical relations between the
different St\"{a}ckel systems. The proposed change of the time is related
to ambiguity of the Abel map. For the two degree of freedom systems that
were studied in this paper, we found the Lax representations and the
$r$-matrix algebras. The corresponding dynamical $r$-matrices have the
intriguing connections to the Drinfeld twists.

Of course, considered above particular family of the time transformations
(\ref{dham}) does not  exhausted all the set of the non-canonical changes
of the time, which preserve the integrability. As an example, the
complete Kolosoff transformation $\{t,\p,\q\}\to\{\te{t},\te{\p},\q\}$
\cite{kol01} connects the St\"{a}ckel system with the other integrable
non-St\"{a}ckel system. So, it would be interesting to investigate
another integrable systems connected with the St\"{a}ckel systems by
non-canonical transformations.

This work was partially supported by RFBR grant.


\end{document}